\documentclass[12pt,a4paper]{article}

\usepackage[preprint,nonatbib]{nips_gibs}

\usepackage[T1]{fontenc}
\usepackage[utf8]{inputenc}

\usepackage{float}
\usepackage{graphicx}
\usepackage{amssymb,amsmath}
\usepackage{hyperref}

\usepackage[export]{adjustbox}

\newcommand{\reflink}[1]{\href{#1}{#1}}

\usepackage{authblk}

\usepackage{enumitem}
\setitemize{noitemsep,topsep=0pt,parsep=2pt,partopsep=2pt}
\setenumerate{noitemsep,topsep=2pt,parsep=2pt,partopsep=2pt}
\setdescription{noitemsep,topsep=2pt,parsep=2pt,partopsep=2pt}

\begin{document}

\title{Computing Accessibility Metrics for Argentina}

\author[1]{Carolina Lang}
\author[1]{Tobias Carreira}
\author[2]{Germán César Dima}
\author[3]{Lucila Berniell}
\author[4]{Carlos Sarraute}

\affil[1]{Computer Science Dept., University of Buenos Aires}
\affil[2]{Ministry of Production and Labor (MPyT), Argentina}
\affil[3]{CAF Development Bank of Latin America}
\affil[4]{Grandata Labs}

\maketitle{}

\begin{abstract}

We present a tool to calculate distances and travel times between a set of origins and a set of destinations, using different modes of transport in Argentina. The input data for the tool is a set of destinations (a geo-referenced list of points of city amenities or ``opportunities'', such as firms, schools, hospitals, parks, banks or retail, etc.) and a set of origins characterized by their geographic coordinates that could be interpreted as households or other. The tool determines, from each origin, which is the closest destination, depending on the distance or travel time and the mode of transport (on foot, by bicycle, by car, and by public transport).

The sets of origins and destinations are large sets, which can contain up to several thousand points. We applied and developed algorithms to improve the scalability of the different parts of the procedure. For the public transportation network, we pre-processed the reachable lines from each point and used quad-trees to determine the distance between said points and the bus line's path.

A second objective of this project was to rely only on open data, such as Open Street Map (OSM) data, together with making this tool open source. Therefore, the successful development and implementation of this tool is potentially beneficial to both public sector agencies as well as NGOs and other civil society organizations that focus their work on the design and implementation of public policies, aimed at improving accessibility in cities as a way to reduce spatial inequalities and social exclusion. 

\end{abstract}

%-----------------------------------------------------

% 1
%!TEX root = paper-eng.tex

%--------------------------------------------------------------------------
\section{Introduction}
%--------------------------------------------------------------------------

%--------------------------------------------------------------------------
\subsection{Problem and context}
%--------------------------------------------------------------------------

Urban sprawl, socioeconomic segregation, and the impacts of investments in infrastructure and transportation networks are issues whose common factor is to affect the dynamics of cities and ultimately the wellbeing of people that reside in urban areas, which in the developing world is the vast majority. These problems are influenced by geographic accessibility, that is, the possibility that all people can easily transit, through the use of existing infrastructure, to destinations of their own interest. In the last decade, there has been a focus on generating measures capable of determining the quality of accessibility to destinations such as health centers, firms and schools, because  all of them play a major role in determining many economic and social outcomes of households.

When estimating accessibility, it is common to use online services such as Google Maps. However, these application servers have multiple disadvantages: they function as black boxes and do not allow users to modify their content (e.g., adding information on streets or addresses), they have limitations when searching for large amounts of data and they depend on internet connection to perform any kind of query. In addition, they have information retention policies, which --in many cases-- may conflict with user confidentiality and with the security of the data used in the search.

We propose the creation of the first open tool capable of providing information about atomistic geographic accessibility in Argentina, completely transparent, free, offline and adaptable to the user's needs. This tool contemplates public transport networks and is flexible to the addition of many sets of destinations (``opportunities'') of interest. In addition, it operates with specific locations (that is, both origins and destinations should include geographic coordinates), so it will function as a building block when constructing indicators at lower granularity levels, such as municipalities, provinces or states. 

With this tool, the user will be able to specify the mode used to calculate accessibility: on foot, by bicycle, by car or by public transport. The program will yield a set of measures (distance and time) which will serve as raw material for the preparation of subsequent more elaborate accessibility metrics. Possible uses of this tool are:

\begin{itemize}
\item the construction of indexes capable of summing up in only one dimension the concept of accessibility from a given set of addresses (homes) to the most relevant (economic and social) opportunities for household members,
\item comparisons of accessibility metrics in one specific domain (e.g., distance to a subset of firms or establishments, for instance to those where average paid wages are above a certain threshold) across different origins, which can take the form of isochrones or other ways to summarize the spatial dispersion in the domain of interest,
\item visualizations of saturation indicators (e.g., heat maps) that properly transmit information about opportunities that are accessible or reachable to households residing in different neighborhoods of a given city, as a way to convey useful information for residential and housing decisions.
\end{itemize}

In all these examples, it is not only necessary to have a tool capable of measuring urban accessibility, but also to have detailed information of both starting points (homes) and arrivals (opportunities). In this aspect, the Ministry of Production and Labor (MPyT) has an outstanding record in the field, handling administrative records as well as surveys and censuses, which allow us to characterize households and to know their potential needs with respect to opportunities. Likewise, in the use cases discussed in Section~\ref{use-cases}, we will make use of some open databases (e.g., schools, hospitals, banks, etc.) that are available nationwide and which contain georeferenced and some other useful descriptive features of these type of destinations.

In addition to the new lines of research opened by this tool, we can list other areas where its application may add a quantitative dimension to the already existing approach --usually qualitative, or quantitative but lacking the spatial dimension behind the policy issue-- for designing and analyzing relevant public policies. For instance, this tool can help to improve the policy discussion around several relevant policy issues: 
	
\begin{itemize}
\item The analysis of spatial labor mismatches (analysis of labor supply and demand) as well as the assessment of how existing metrics of accessibility correlate with the degree of misallocation of resources induced by such mismatches.
\item The analysis of commuting time costs, which is usually done by means of (sometimes quite costly) origin-destination surveys which serve as the basis for the design, implementation and evaluation of policies regarding urban transport planning.
\item The analysis of place-based policies, which target places --instead of individuals or firms-- and try to boost economic activity in relatively narrowly defined geographic areas (e.g., enterprise zones).
\item The study of individuals' and families' residential and housing decisions (human mobility and migrations).
\end{itemize}

%--------------------------------------------------------------------------
\subsection{Proposed data science product}
%--------------------------------------------------------------------------

The objective of the project is to develop a tool to calculate distances and travel times between:
\begin {itemize}
\item a set of origins,
\item a set of destinations,
\item using different modes of transport.
\end {itemize}

The input data (inputs) for the tool are:
\begin {itemize}
\item Destinations: A geo-referenced list of points of interest or opportunities (such as employers' establishments, schools, banks, libraries, etc.)
For example: a list with over 50,000 schools in Argentina.

\item Origins: The tool includes several ways to specify the origins:
\begin {itemize}
\item List of points (geographic coordinates).
\item For example: coordinates of residences of households surveyed in official large-scale household surveys.
\item All the corners in a given urban area (e.g., the Autonomous City of Buenos Aires, CABA, or the Metropolitan Area of Buenos Aires, also known as AMBA).
\item Centroids of census tracts.
\item Centroids of polygons defining legal limits for municipalities.
\end{itemize}

\end{itemize}

Based on the input data, the tool is responsible for determining, from each origin, which is the closest destination, depending on the distance or travel time and the mode of transport. The options that we contemplate are:
\begin {itemize}
\item Minimum distance:
\begin {itemize}
\item On foot
\item By car

\end {itemize}

\item Minimum travel time:
\begin {itemize}
\item On foot
\item By bicycle
\item By car
\item By public transport.
\end{itemize}

\end{itemize}

It is worth mentioning that the sets of origins and destinations are large sets, since each one can contain up to several thousand points. This makes calculating the minimum path from all origins to all destinations a task that is computationally very inefficient. To overcome this problem, we developed a considerably more efficient implementation that calculates minimum paths.

A second objective of this project was to rely only on open data, that is, to use freely available and freely accessible data, such as Open Street Map (OSM) data. We consider that using open data, together with making this an open source tool, are crucial aspects to not depend on external factors such as the data of a private company or the payed access to an API of a private company.

%--------------------------------------------------------------------------
\subsection{Related work}
%--------------------------------------------------------------------------

Previous works have recognized not only the importance of using quantitative measures of accessibility in the design of urban policies, but have also shown the inherent complexity of constructing such metrics, specially in the context of developing countries~\cite{red2017}. However, and for the specific case of the city of Buenos Aires, some of the collaborators on this project developed interesting metrics that partially capture the idea of accessibility and that made use of a completely different type of data (cell phone data)~\cite{Anapolsky2014exploracion,Sarraute2017city}. It is worth mentioning that both the approach taken here and the approach in~\cite{Anapolsky2014exploracion,Sarraute2017city} are perfectible compatible and are likely to enrich one another. Some other related work can be found in 
studies on the prediction of human mobility, for instance those that include the analysis of particularly large sociocultural events~\cite{Ponieman2013human,Ponieman2015mobility}, or the analysis of more regular human mobility patterns~\cite{Mucelli2016regularity}, and even the  measurement-driven mobile data traffic modeling~\cite{Oliveira2015measurement}.

% 2
%!TEX root = paper-eng.tex

%-----------------------------------------------------
\section{Data sources}
%-----------------------------------------------------

%-----------------------------------------------------
\subsection{Open Street Map data}
%-----------------------------------------------------

As previously stated, we propose to use data from \textit{Open Street Map} (OSM), which can be downloaded and used freely~\cite{OpenStreetMap}.
OSM is a public repository of geographic information whose data quality has established it as a frequent source for mobility studies~\cite{haklay2010good,juran2018geospatial}.

OSM has a repository of data from around the world, including different establishments and services. Its main data are the streets of different cities,
which is what interests us the most for the present work.
In addition, as it is an open and collaborative project, it has a large community that continuously improves the data and develops different tools, such as the \emph{Open Source Routing Machine} (OSRM) described in Section~\ref{sec:osrm}.

All data are modeled as \textit{nodes}, \textit{ways} or \textit{relations}.
The nodes represent point objects in space, defined by their latitude, longitude and \textit{id}.
They represent streets, shops and other establishments.
Roads are defined as ordered lists of points.
They are used to model streets, routes and areas (such as parks).
Finally, relationships are groups of other elements.
For example, relationships can be used to model cities, provinces or countries,
grouping all their streets, establishments and other elements of interest.

To prevent our work from requiring access to the internet,
we decided to download the data corresponding to a sector of our interest
using the API \textit{Overpass}\footnote{https://wiki.openstreetmap.org/wiki/Overpass\_API},
which is designed and optimized for this use.
Once downloaded, we can make use of this data freely.

%-----------------------------------------------------
\subsection{Public Transportation Data of CABA}
%-----------------------------------------------------

The City of Buenos Aires has a large repository of open \textit{datasets}~\cite{bsas2019}.
In particular, we are interested in using public transport data,
to be able to compute accessibility measures by bus.

Both buses and subway \textit{datasets} have schedules in the format
given by the General Transit Feed Specification
(GTFS)~\cite{gtfs2019}.
The bus routes are available for all the lines. However, the bus timetable data are available for only 16 lines.
Because of this, we are faced with the challenge of trying to approximate the time of the rest of the lines (wherein the bus stops data are found, but without timetables), as we explain later.

% 3
%!TEX root = paper-eng.tex

%-----------------------------------------------------
\section{Calculation of distances by car / bicycle / on foot}
%-----------------------------------------------------

%-----------------------------------------------------
\subsection{Use of OSRM to calculate distances between points}
\label{sec:osrm}
%-----------------------------------------------------

As previously mentioned, the OSM community generated different tools around the data repository.
One of them is the \textit{Open Source Routing Machine} (OSRM), which is used to calculate roads, distances and travel times using different means of transport: on foot, by bicycle or by car~\cite{huber2016calculate}.
Given the stability and optimization of this tool, we prefer to use it over doing our own routing.

In addition, it has different possible services, from which we highlight the \textit{Table} mode, which allows the calculation of travel times and distances between a set of origins and one of the destinations, and the \textit{Route} mode, which allows us to do the same given an ordered list of points.

OSRM can be used in different ways:
\begin{enumerate}
\item Through its HTTP API, using available servers.
\item Through its HTTP API, building our own local web server.
\item As a C ++ library.
\end{enumerate}

We decided to go for the third option, since it does not require an internet connection and is faster to use than setting up our own local web server (since HTTP overhead is avoided).

\begin{table}[h]
\centering
	\caption{OSRM Extract Compression Ratio}
	\begin{tabular}{ |l|l|l|l| }
		\hline
								& Car		& Bicycle	& Foot \\
		\hline
		Node compression ratio	& 0.522412 	& 0.542501	& 0.54018 \\
		Edge compression ratio	& 0.648841 	& 0.655234	& 0.653472 \\
		\hline
	\end{tabular}
	\label{tab:osrm-compression-ratio}
\end{table}

OSRM comes with tools to preprocess OSM data in order to optimize it,
extracting the most useful data and compressing it to save memory space.
As we can see in Table˜\ref{tab:osrm-compression-ratio},
the compression ratio goes from 0.52 to 0.55 for the nodes
and from 0.64 to 0.66 for the edges.

%-----------------------------------------------------
\subsection{Computing the minimum distance to a set of points}
%-----------------------------------------------------

In order to efficiently calculate the distance to the nearest point, with respect to each point $ p $ in the input set, an approach based on $ K $-nearest neighbors ($ K $-NN) was used.

The premise is that calculating the geodesic distance between two points is fast --in particular it is O(1)-- while calculating the distance in the transportation grid has an algorithmic complexity at least proportional to the size of the street graph.

In order to give an efficient solution, we assume that with very high probability, the point of interest closest to $ p $ in the grid of streets will be found among the $ K $ nearest neighbors of $ p $ measuring the geodesic distance, for some $ K $ large enough, but which in turn is considerably smaller than the total number of points of interest.

Therefore, for each point, we proceed as follows:
\begin{enumerate}
\item The geodesic distance (``as the crow flies'') is calculated for each of the points of interest.
\item Only the $ K $ candidates that are closer to $ p $ are selected based on the calculated distance.
\item We calculate for the $ K $ points, using OSRM, the distance to $ p $ according to the streets network, and according to the mode of transport chosen.
\item The lowest value is taken and returned as a result of the algorithm.
\end{enumerate}

The program that executes this part takes as parameters the file of initial points, the file of points of interest, the $ K $ value to calculate $ K $-NN, and the mode it should use for the calculation of distances (on foot, by car or bicycle).

The program currently prints the shortest time needed to reach a point of interest, but it can be easily modified to print the smallest distance according to the street grid, the nearest point or some combination of this information.

% 4
%!TEX root = paper-eng.tex

%--------------------------------------------------------------------------
\section{Calculation of distances and times using public transport}
%--------------------------------------------------------------------------

\subsection{Estimate schedules for lines that do not have schedules grid}

Having only the schedule of a very limited subset of bus lines, we decided to estimate the duration of the trips using OSRM.
For that, we used the method \textit{Route} with the points of all the stops of each line, simulating the route as if it were in automobile.
Counting all branches and directions, we processed 1,113 different lines, where each line has 464 stops on average.

\subsection{Search for paths and stops between source disk and destination disk}

To find which bus lines can be reached from a given origin point, we set a limit $ L $, in meters, and we consider that all bus lines that have a stop at a distance smaller than $ L $ are reachable (considering the geodesic distance).
The algorithm processes each entry point and returns, for each reachable line, the closest stop to the point (again using geodesic distance).

Internally, the bus data and the points to be processed are parsed in the file, which internally invokes the \textit{min\_distances} procedure.
Within this file, a preprocessing of the bus lines is done to make the search more efficient, since measuring the distance between pairs of points for each entry point with respect to each stop of each bus line is too expensive in terms of running time.

The preprocessing consists of saving the bus stops in a \textit{quaternary tree}\footnote{\reflink{https://en.wikipedia.org/wiki/Quadtree}}, slightly modified so that it returns the point $ (0, 0) $ if the distance between all points of a certain portion of the bus line is farther than $ L $.

To measure the geodesic distance between two points, we used the formula of \textit {haversine}\footnote{\reflink{https://es.wikipedia.org/wiki/F\%C3\%B3rmula\_del\_haversine}}, a standard formula that assumes that the Earth is a sphere to calculate distances, which in an area of the planet as small as the City of Buenos Aires is sufficient.

The theoretical complexity is to process, for each point and for each line, the corresponding quaternary tree, in logarithmic time with respect to the size of the tree, so the complexity is:
$$O\left( | P | \cdot \sum_{i=1}^{| lines |} ln(| lines[i] |) \right)$$
where $ lines $ is the list of bus lines, which in each index has the list of stops of that line, and $ P $ is the set of points from which you want to obtain the closest stops of the reachable lines.

\subsection{Adaptation of $ K $-NN for public transport mode}

To calculate the trips in public transport between two points $ P $ and $ Q $, we consider, due to technical limitations, only trips made in one bus without transfers.

Remember that the bus lines are separated according to branches and direction. Therefore, in the context of this work, when we speak of a \textit{bus line}, we are really talking about a directed path (that is, one can traverse it in one direction, and not in the other).

If a trip between $ P $ and $ Q $ uses the $ L $ line, the stages of the trip are:\begin{enumerate}
\item Walk from $ P $ to some stop $ P_L $ of $ L $.
\item Use the bus line from this stop to another stop $ Q_L $.
\item Walk from $ Q_L $ to $ Q $.
\end{enumerate}
and the total time of this particular journey is the sum of the times obtained for these three stages. Waiting times at the bus stop are not considered.

For this trip to be feasible, $ P_L $ must be reachable from $ P $, $ Q $ from $ Q_L $, and $ Q_L $ must be after $ P_L $ in the bus route.

For the calculation of $ K $-NN using public transport, the points of origin and destination must contain the field \textit{stops}, with the stops of the lines reachable from that point.

In addition, the pre-estimate of schedules is used for stops along the route of the bus to create a structure that consults the schedule as an additive table:\begin{verbatim}
busTravelTime(ori, dest, bs):
  time <- bs.timetable[dest] - bs.timetable[ori]
  if time < 0:
    return INF
  return time
\end{verbatim}
where \texttt{INF} is a very large number that represents infinity, and \texttt{bs} is a map that contains the information of the buses.

If the travel time is negative, it means that you are trying to travel in the opposite direction, and therefore you have to discard the trip.

The minimum route is obtained by testing the route for all the lines that are reachable from $ P $ and from $ Q $. In case there is no line that coincides for both points, the algorithm returns the time it takes to walk from $ P $ to $ Q $ according to OSRM.

% 5
%!TEX root = paper-arXiv.tex

%--------------------------------------------------------------------------
\section{Use Cases}
\label{use-cases}
%--------------------------------------------------------------------------

In this section we show the result of applying the algorithm, for the different modes of transport,
using origin points that were sampled randomly within each district of the city of Buenos Aires (CABA), and distributed according to the population of the districts. The city is divided in 48 districts. We used the information from the 2010 census\footnote{Data made available by General Direction of Statistics and Census (Ministry of Economy of the City of Buenos Aires), based on the INDEC database of the 2010 National Census. Available at \url{https://www.estadisticaciudad.gob.ar/eyc/?p=28011}}.

The sets of points that we use as destinations are schools, banks, security forces and public hospitals.

%-----------------------------------------------------
\subsection{Schools}
%-----------------------------------------------------

\begin{figure}[th]
\centering
\includegraphics[width=.52\linewidth,valign=t]{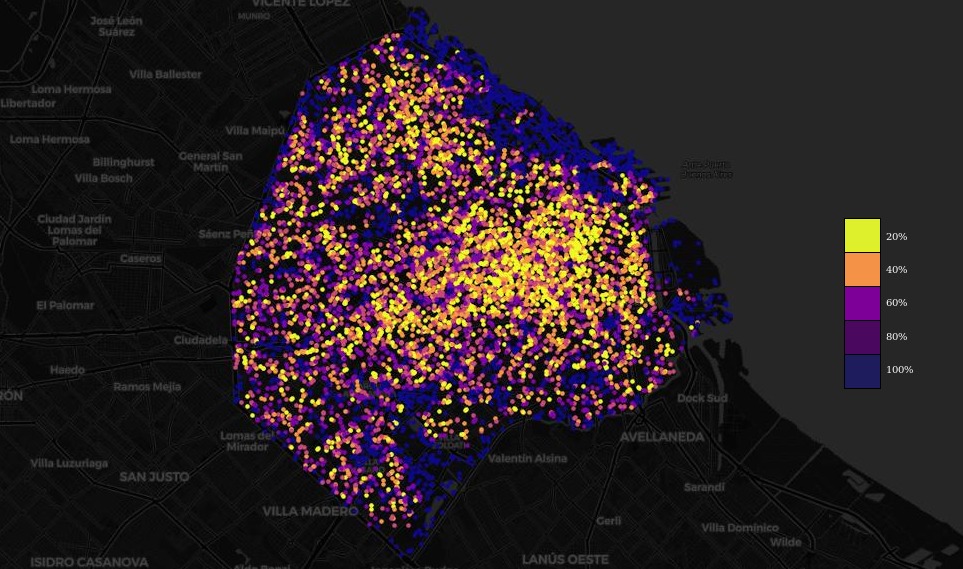}	
\includegraphics[width=.44\linewidth,valign=t]{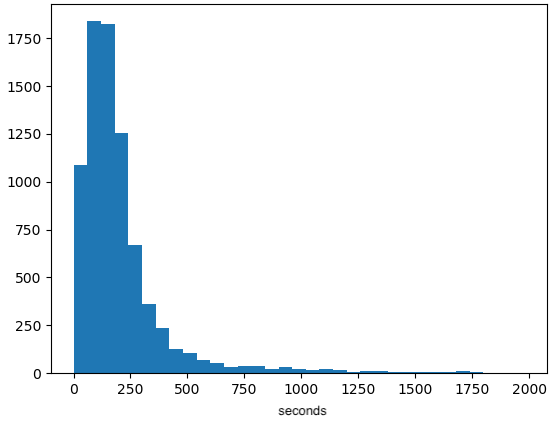} 
\caption{Schools -- Heatmap and distribution of travel time (in seconds) on foot.}
\label{fig:escuelas-pie}
\end{figure}

Figure~\ref{fig:escuelas-pie} shows a heat map of the times of access to schools in CABA on foot, where the set of origins consists of 8,000 points sampled according to the population density, and the set of destinations contains 22,102 schools. The colors of the heat map correspond to the quantiles of travel times, where yellow indicates the shortest travel time quantiles, and violet indicates the longest travel time quantiles. The figure also displays the distribution of travel times, measured in seconds.

\begin{figure}[th]
\centering
\includegraphics[width=.52\linewidth,valign=t]{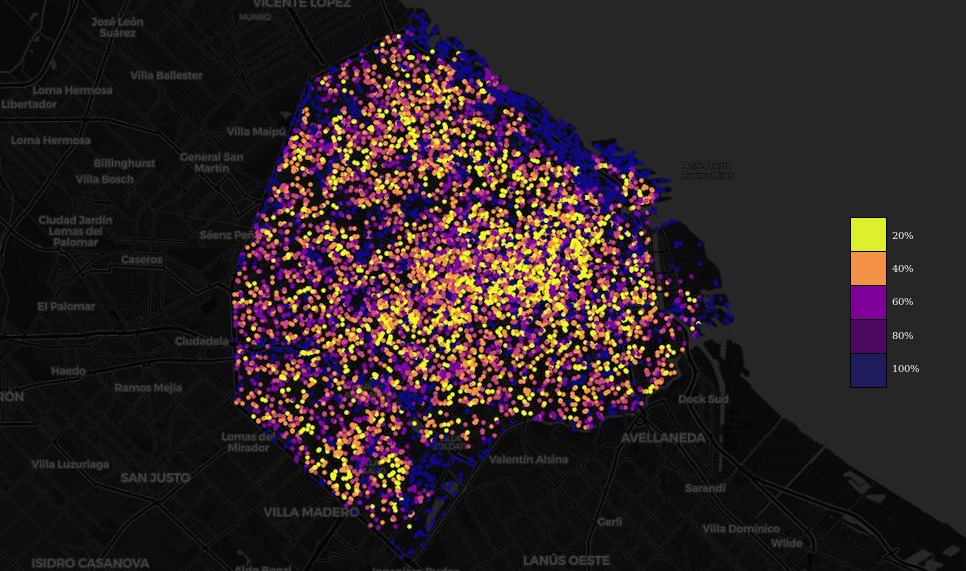}
\includegraphics[width=.44\linewidth,valign=t]{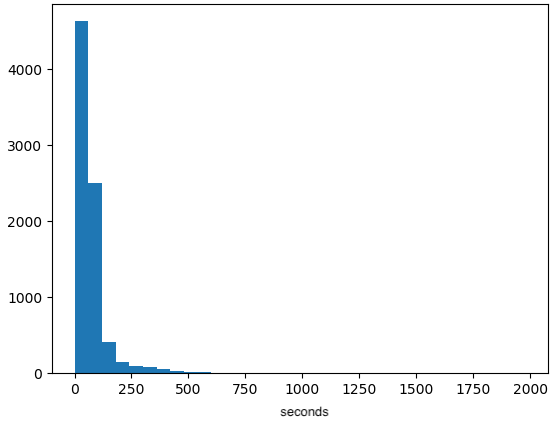} 
\caption{Schools -- Heatmap and distribution of travel time (in seconds) by car}
\label{fig:escuelas-auto}
\end{figure}

\begin{figure}[th]
\centering
\includegraphics[width=.52\linewidth,valign=t]{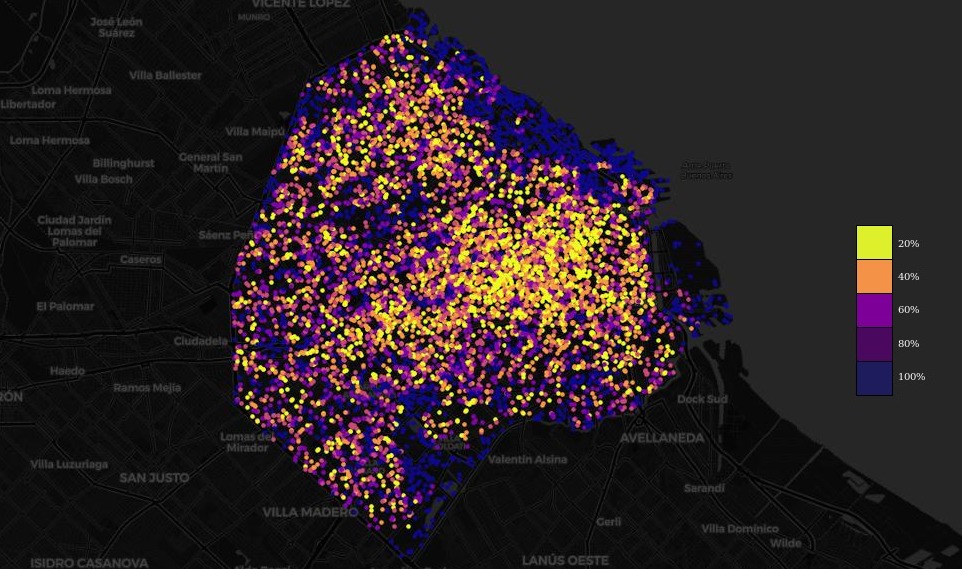}
\includegraphics[width=.44\linewidth,valign=t]{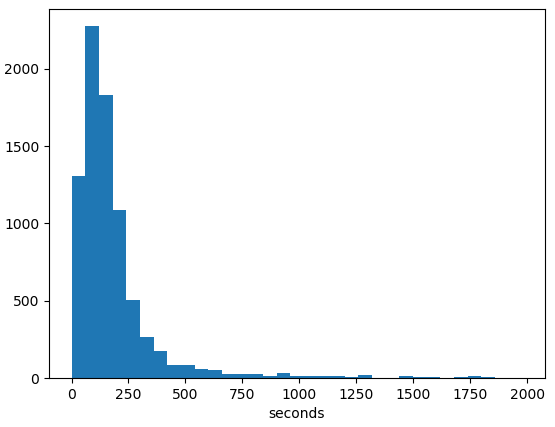} 
\caption{Schools -- Heatmap and distribution of travel time (in seconds) by public transport}
\label{fig:escuelas-tp}
\end{figure}

Figures~\ref{fig:escuelas-auto} and \ref{fig:escuelas-tp} show the heat map and the distribution of travel times to schools by car and by public transport (by bus taking $L = 500 \,m$ as the limit for the walking distance to the bus stop), with the same number of origins. We can observe that the number of schools and their distribution in the city makes them accesible from all the neighborhoods, using any mode of transport.

%-----------------------------------------------------
\subsection{Security forces}
%-----------------------------------------------------

\begin{figure}[h!]
\centering
\includegraphics[width=.52\linewidth,valign=t]{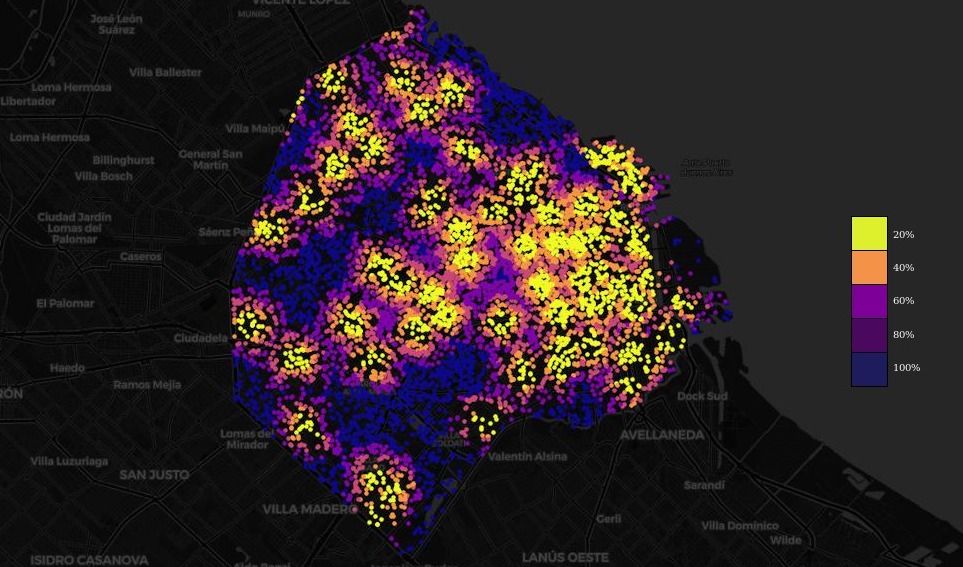}
\includegraphics[width=.44\linewidth,valign=t]{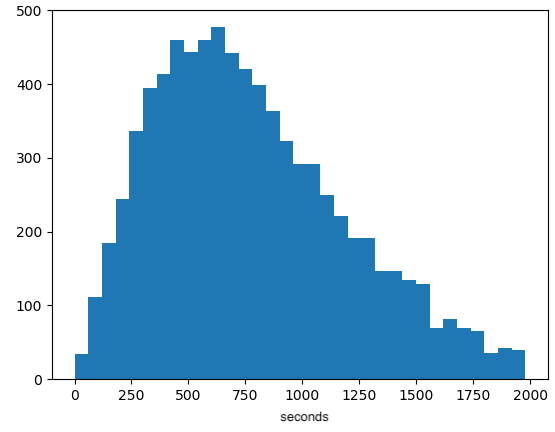} 
\caption{Security forces -- Heatmap and distribution of travel time (in seconds) on foot}
\label{fig:ffaa-pie}
\end{figure}

Figure~\ref{fig:ffaa-pie}  shows a heat map and the distribution of the times of access to security forces in CABA on foot, using a set of 8,000 origin points sampled according to the population density of the districts. The destinations consist of 292 security forces' establishments.

There is a clear contrast with the times of access to schools: the number of security forces establishments is much smaller, and are more concentrated in the downtown and the city center. There are large sectors of the city that require longer walking times to reach this type of establishments.

\begin{figure}[h!]
\centering
\includegraphics[width=.52\linewidth,valign=t]{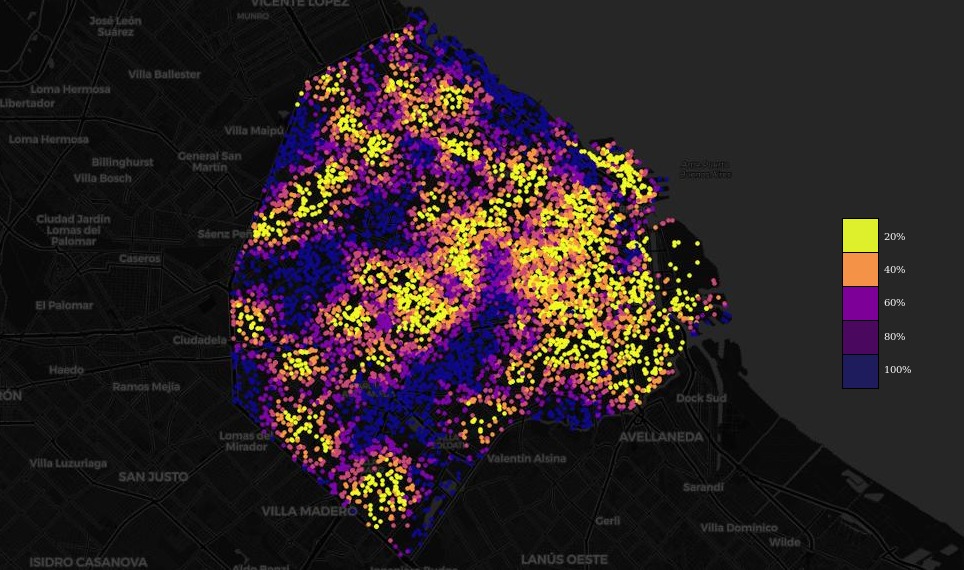}
\includegraphics[width=.44\linewidth,valign=t]{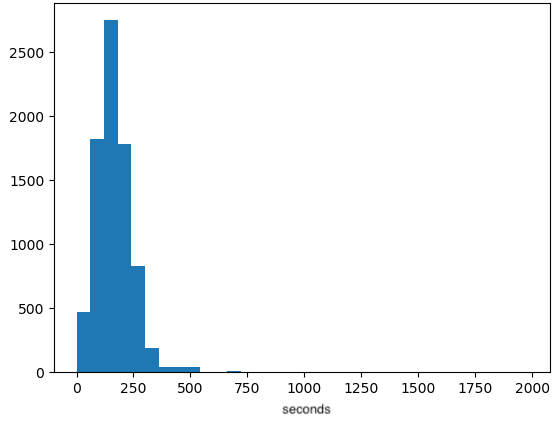} 
\caption{Security forces -- Heatmap and distribution of travel time (in seconds) by car}
\label{fig:ffaa-auto}
\end{figure}

\begin{figure}[h!]
\centering
\includegraphics[width=.52\linewidth,valign=t]{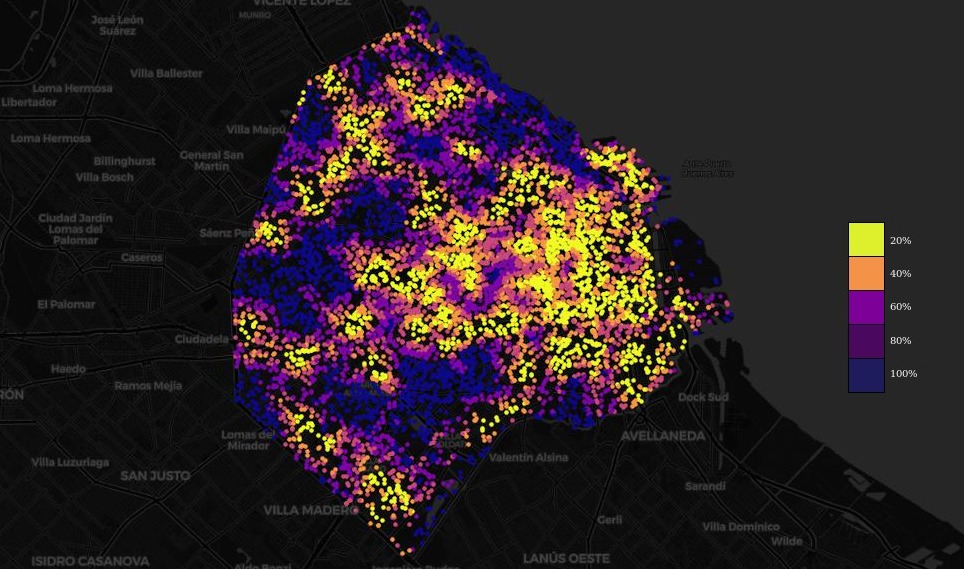}
\includegraphics[width=.44\linewidth,valign=t]{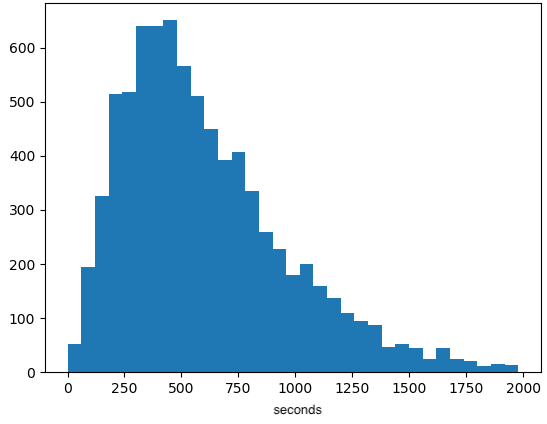} 
\caption{Security forces -- Heatmap and distribution of travel time (in seconds) by public transport}
\label{fig:ffaa-tp}
\end{figure}

Figures~\ref{fig:ffaa-auto} and \ref{fig:ffaa-tp} show the heat map of access to security forces by car and by public transport (bus).
We can observe here the contrast between the two transport modes: accessibility by car is very efficient, whereas access by bus takes a much longer time in our estimation.

%-----------------------------------------------------
\subsection{Banks}
%-----------------------------------------------------

\begin{figure}[h!]
\centering
\includegraphics[width=.52\linewidth,valign=t]{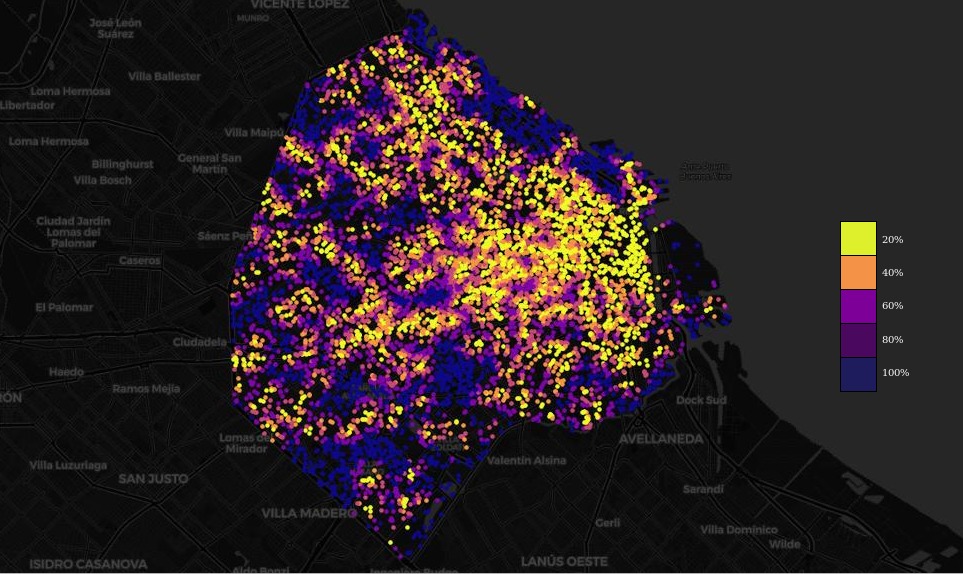}
\includegraphics[width=.44\linewidth,valign=t]{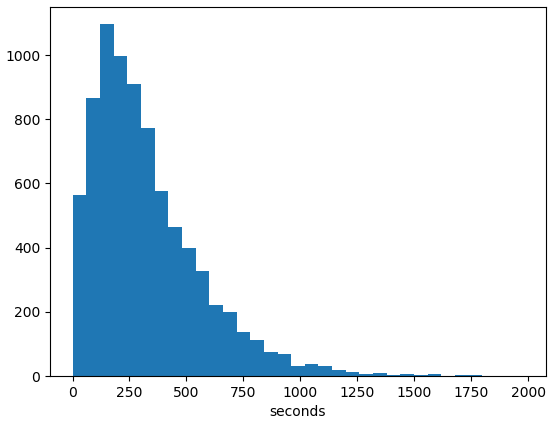} 
\caption{Banks -- Heatmap and distribution of travel time (in seconds) on foot}
\label{fig:bancos-pie}
\end{figure}

Figure~\ref{fig:bancos-pie} shows the heat map and distribution of the access times to banks in CABA on foot (measured in seconds). The origins are 8,000 points sampled randomly according to the population density of the neighborhoods of the city. The destinations are 4,202 banks.

\begin{figure}[h!]
\centering
\includegraphics[width=.52\linewidth,valign=t]{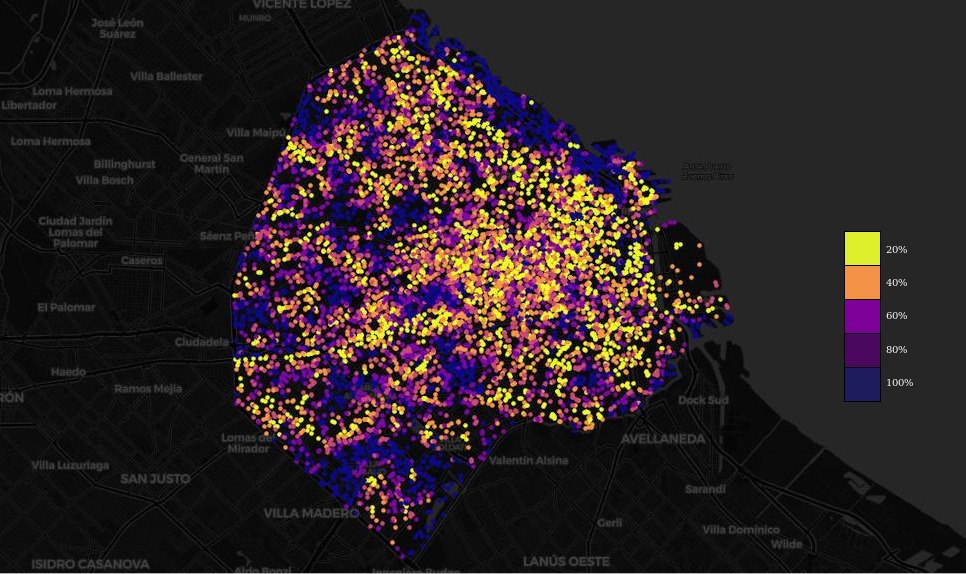}
\includegraphics[width=.44\linewidth,valign=t]{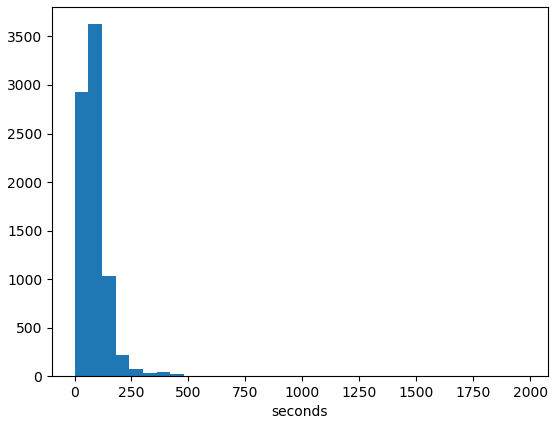}
\caption{Banks -- Heatmap and distribution of travel time (in seconds) by car}
\label{fig:bancos-auto}
\end{figure}

\begin{figure}[h!]
\centering
\includegraphics[width=.52\linewidth,valign=t]{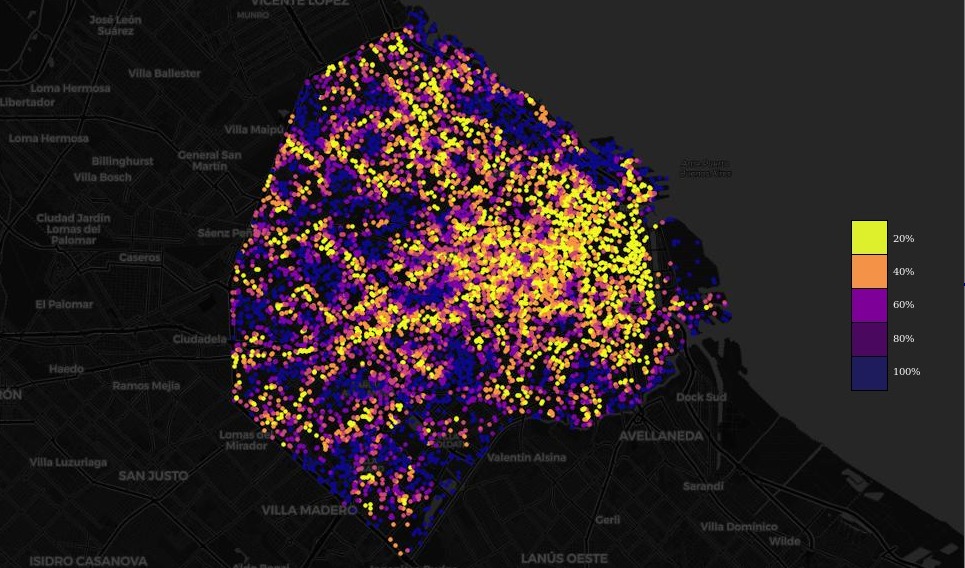}
\includegraphics[width=.44\linewidth,valign=t]{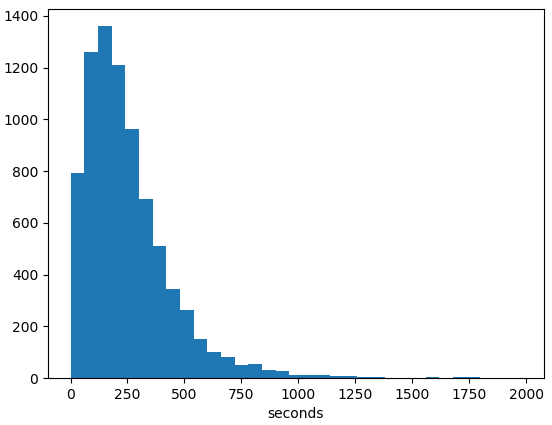}
\caption{Banks-- Heatmap and distribution of travel time (in seconds) by public transport}
\label{fig:bancos-tp}
\end{figure}

Figure~\ref{fig:bancos-auto} shows the heat map and distribution of access times to banks by car. This situation of the banks is an intermediate between the distribution of schools and security forces. The banks have a very good coverage of the whole city, and are readily accessible from the whole city by car. However, for the citizens going to the bank on foot, some neighborhoods have longer traveling times. Figure~\ref{fig:bancos-tp} displays the heat map and distribution of access times to banks by public transport, showing that banks also have a good coverage for citizens traveling by bus.

%-----------------------------------------------------
\subsection{Hospitals and Health Centers}
%-----------------------------------------------------

We had access to a dataset of Health Providers, with location records of the greatest possible number of health providers in Argentina, located with latitude and longitude coordinates.
This dataset was built by Antonio Vázquez Brust, Tomás Olego and Germán Rosati from Fundación Bunge y Born by integrating different sources of official data:
National Base of Hospitals and Primary Care Centers compiled by the Argentine Health Information System~\cite{sisa2019};
SUMAR program health providers~\cite{sumar2019};
and a list of hospitals and health care centers from the
National Program for Sexual Health and Responsible Procreation
(Ministry of Health)~\cite{msal2019}.

As part of the dataset cleaning process, the health providers were classified according to their level of complexity. 
There were different classifications in the datasets used related to the notion of complexity, and the criteria varied among sources: the classifications were not homogeneous in the different lists of health providers consulted.
A classification that unifies the different denominations
was produced, resulting in a simple classification into three categories, in decreasing order of complexity:
Hospital, Health Center and Sanitary Post.
After cleaning and classifying, 13,910 records were obtained for the whole country.

There are only 110 records that belong to the studied area (city of Buenos Aires).
There is such a small number of public institutions because a large proportion of healthcare establishments in the capital city are private, and we do not consider private providers in this study.

\begin{table}[ht]
\centering
\caption{Healthcare type distribution in Buenos Aires city}
\begin{tabular}{| p{0.45\linewidth} | c | c | c | c |}
\hline
Type of coverage/Year					& 2017	& 2016	& 2015	& 2014	\\
\hline
Public healthcare						& 18.7\%	& 20.0\%	& 18.7\%	& 17.8\% \\
Work health insurance				& 46.1\%	& 40.6\%	& 42.1\%	& 44.4\%	\\
Work health insurance (outsourced)	& 17.3\%	& 19.9\%	& 21.6\%	& 20.8\%	\\
Work health insurance (total)		& 63.4\%	& 60.5\%	& 63.7\%	& 65.2\%	\\
Other private healthcare				& 17.9\%	& 19.4\%	& 17.6\%	& 17.1\%	\\
\hline
\end{tabular}
\label{tab:healthcare-type}
\end{table}

Table \ref{tab:healthcare-type}
 shows the distribution of the healthcare type in the city in the past few years\footnote{\reflink{https://www.estadisticaciudad.gob.ar/si/genero/principal-indicador?indicador=tip\_cob\_sal}}.
We can see that most of the people receive healthcare insurance as a part of their formal work contract. This corresponds to most of the city's formal workforce.
The private healthcare has a high correlation to a high social status, while the more vulnerable population (unemployed and informal workforce) is the main user of the public healthcare system.

This means that, although we lack information about many of the healthcare institutions in the city, the public ones are the most interesting ones if we can combine them with another accessibility measures like the Socio-Economic Index (SEI), because the datum by itself can be misleading when drawing conclusions about areas with a high SEI.

\begin{figure}[h!]
\centering
\includegraphics[width=.52\linewidth,valign=t]{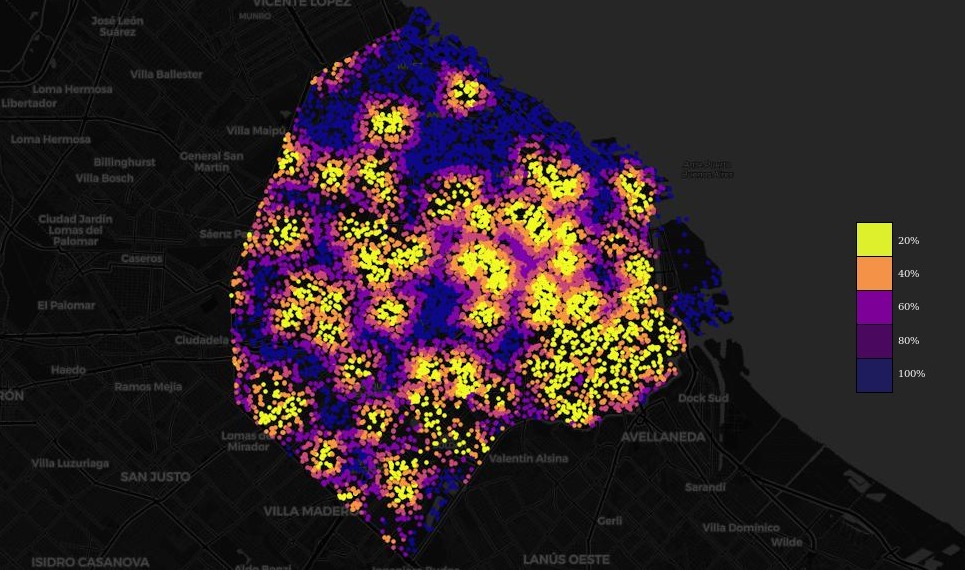}
\includegraphics[width=.44\linewidth,valign=t]{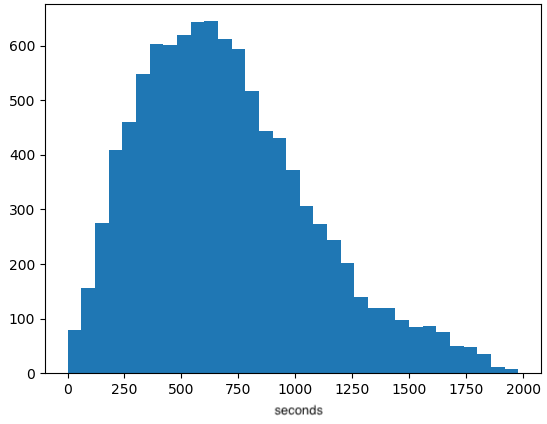} 
\caption{Hospitals -- Heatmap and distribution of travel time (in seconds) on foot}
\label{fig:hosp-pie}
\end{figure}

\begin{figure}[h!]
\centering
\includegraphics[width=.52\linewidth,valign=t]{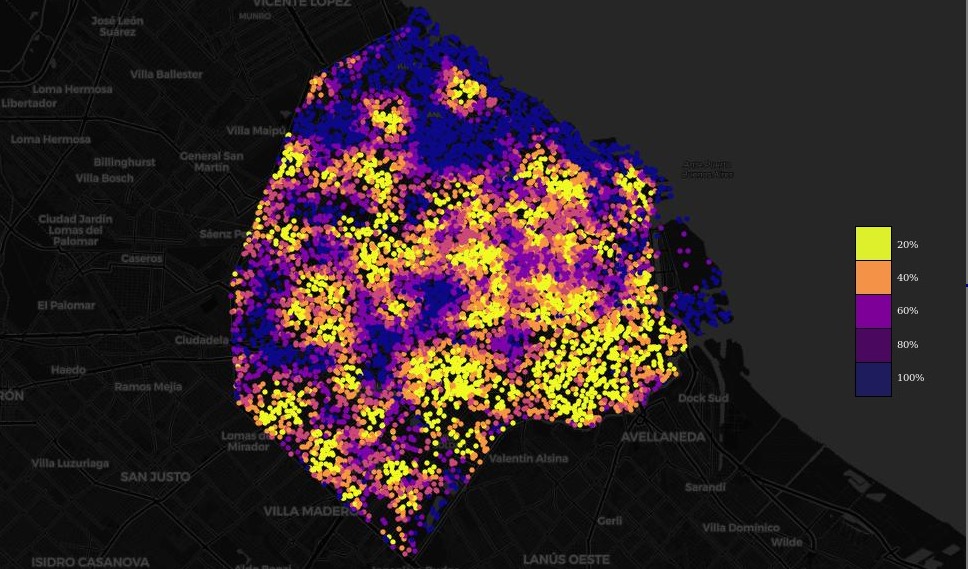}
\includegraphics[width=.44\linewidth,valign=t]{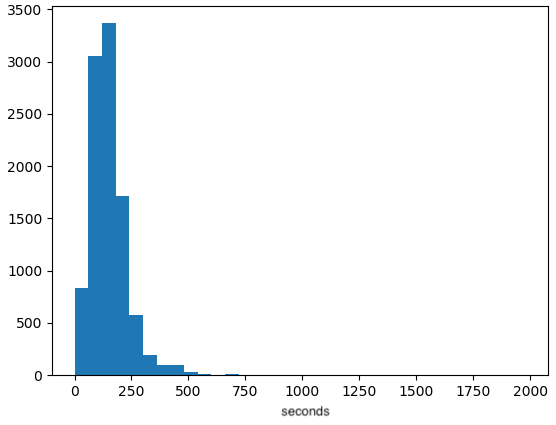} 
\caption{Hospitals -- Heatmap and distribution of travel time (in seconds) by car}
\label{fig:hosp-auto}
\end{figure}

\begin{figure}[h!]
\centering
\includegraphics[width=.52\linewidth,valign=t]{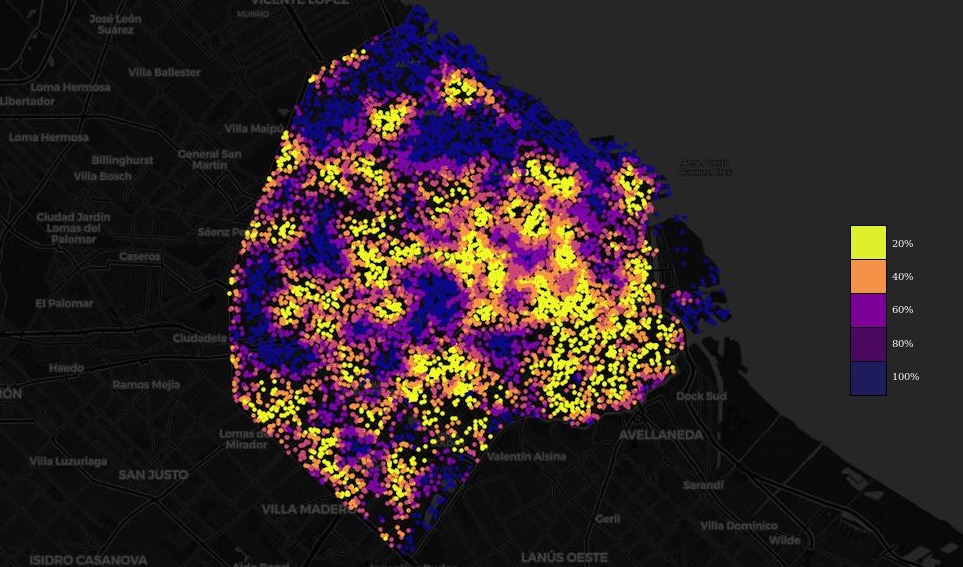}
\includegraphics[width=.44\linewidth,valign=t]{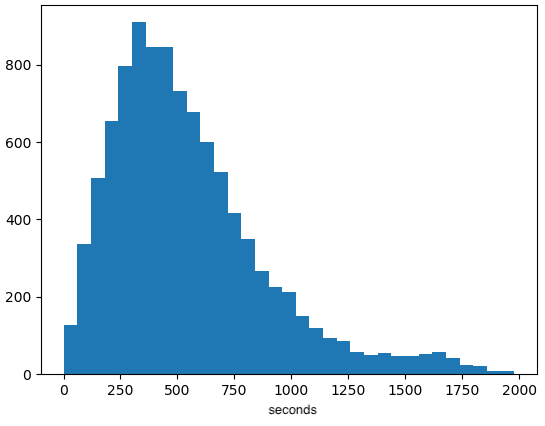} 
\caption{Hospitals-- Heatmap and distribution of travel time (in seconds) by public transport}
\label{fig:hosp-tp}
\end{figure}

Figures~\ref{fig:hosp-pie} and \ref{fig:hosp-auto} show heat maps and distributions of the access times to public health providers in CABA on foot and by car. Figure~\ref{fig:hosp-tp} shows the heat map of access to public health providers by public transport (by bus).

We can observe that the northern part of the city, of higher socio-economic status, has a lower coverage of public health providers inside the city limits. This is coherent with the information of Table~\ref{tab:healthcare-type}, since the health coverage in the wealthier parts of the city is provided by the private health system. Another possibility for those citizens is to travel to health centers in the suburbs.
As a result, citizens requiring access to public health providers have longer travel times on foot and by public transport.

% 6
%!TEX root = paper-eng.tex

%--------------------------------------------------------------------------
\section{Conclusion and future work}
%--------------------------------------------------------------------------

This project was a joint work between the Ministry of Production and Labor (MPyT) of Argentina, CAF Development Bank of Latin America and Grandata Labs. One of our challenges was the collaboration between actors from different sectors (public and private) and different backgrounds (economy, urbanism, physics, mathematics and computer science). 

We developed a tool to calculate efficiently optimal distances and travel times between sets of origins and destinations, using different modes of transport, wherein the sets of origins and destinations can contain several thousands points. For this development we relied exclusively on open data, and will make the resulting code open source. We also tested this tool on use cases relevant for the public sector.

We believe the successful implementation of this tool can be of great help for both public sector agencies as well as NGOs and other civil society organizations that focus their work in the design and implementation of public policies, aimed at improving accessibility in cities as a way to reduce spatial inequalities and social exclusion. 

\smallskip
These are some possible continuations of this project:
\begin{itemize}
\item Several optimizations of the running time:
	\begin{itemize}
\item Parallelization of several stages of the process.
\item Use of quaternary trees for the set of points of origin and destination.
	\end{itemize}
\item Refine the parameters:
	\begin{itemize}
\item Calculation of the time multiplier by bus with respect to time in car.
\item Set the number of candidate points of $ k $-NN.
\item Evaluate the redefinition of the acceptable walking distance to / from a bus stop.
	\end{itemize}
\item Add subway and train lines to the algorithm.
\item Apply to new geographic areas.
\end{itemize}

These are additional ideas that emerged during the final project meeting:\begin{itemize}

\item Compare the times and trajectories with those returned by the Google Maps API (for a reasonable sample).
\item Visualize the routes on the map.
\item Visualize the use of the traffic network.
\item Allow as input a set of polygons (for example neighborhoods or census tracts) with a number of points to be sampled within each polygon.
\item Add a cost function per person modeled, including the value of tickets, and the valuation of travel time per person.
\item For public transport: experiment with different walking distances to the bus stop.

\end{itemize}

%-----------------------------------------------------
% References
%-----------------------------------------------------
\bibliographystyle{unsrt}

\bibliography{../bibliography/mobility}{}
%-----------------------------------------------------

\end{document}